\documentclass[11pt,twoside]{article}
\usepackage{./asp2014}

\aspSuppressVolSlug
\resetcounters

\begin{document}

\title{The Research Data Alliance}
\author{Fran\c coise Genova,$^1$ 
\affil{$^1$CDS, Observatoire astronomique de Strasbourg, Universit\'e de Strasbourg, CNRS, UMR 7550, 11 rue de l'Universit\'e, F-67000, Strasbourg, France; \email{francoise.genova@astro.unistra.fr}}}

\paperauthor{Francoise Genova}{francoise.genova@astro.unistra.fr}{ORCID_Or_Blank}{Author1 CNRS/Université de Strasbourg}{Observatoire Astronomique de Strasbourg}{Strasbourg}{N/A}{F-67000}{France}

\begin{abstract}
The Research Data Alliance (RDA, \url{https://www.rd-alliance.org/}) aims at enabling research data sharing without barriers. It was founded in March 2013 by the Australian Government, the European Commission, and the USA NSF and NIST. It is a bottom-up organisation which after 2 years and a half of existence gathers around 3,000 members from 100 different countries. Work in RDA is organised in a bottom-up way, through Working Groups and Interest Groups proposed by the community. These Groups can deal with any aspect of research data sharing, which means a huge diversity in the activities. Some scientific communities use the RDA as a neutral place to hold the discussions about their disciplinary interoperability framework. Astronomy has the IVOA and the FITS Committee for that purpose, and the ADASS Colloquia to deal with astronomical data systems. But many RDA topics are of interest for us, for instance data citation, including citation of dynamic data bases and data repositories, or certification of data repositories. Also lessons learnt in building the IVOA and data sharing in astronomy are injected in the discussion of RDA organisation and procedures. The RDA is a unique platform to meet data practitioners from many countries, disciplines and profiles, to grasp ideas to improve our data practices, to identify topics of common interest and to raise interesting subjects for discussion involving specialists from many disciplines and countries by proposing new Interest and Working Groups.\end{abstract}

\section{The Research Data Alliance}

The Research Data Alliance high level aim is to build the social and technical bridges that enable open sharing of scientific data. The RDA vision is researchers and innovators openly sharing data across technologies, disciplines, and countries to address the grand challenges of society. Astronomy has been a pionneer and remains at the forefront for the sharing of data, but we know that the scientific data landscape is highly fragmented. The RDA tackles "building blocks" of common data infrastructures and builds specific "data bridges" for enabling data sharing.

The RDA is a bottom-up organisation: it relies on its community to define and perform its activities, which are organised in Working Groups and Interest Groups. The questions relevant to scientific data sharing are very diverse, and the Group activities cover a wide range of subjects, from very technical to "sociological" ones. The outcomes of their activities are also very diverse, specific tools or pieces of codes, but also best practices, standards (although the RDA is not a "standards organisation" like the W3C or the IETF), but also surveys or reports. Groups should have international membership.  Proposals for new Groups are examined by the RDA Technical Advisory Board, an elected body which supervizes the technical activites of the RDA, and by the RDA Council. The whole RDA community is associated to the RDA process, through Request for Comments on proposed Working Groups and Interest Groups and on Group outcomes.

\section{Working Groups and Interest Groups in the RDA}

\textit{Working Groups} are engaged in creating \underline{deliverables} that will directly enable data sharing, exchange, or interoperability. They conduct short-lived, 18 month efforts to produce "implementable" delivrables,  which include, but are not limited to, technical specifications and implementation practices, conceptual models or frameworks, implemented policies, and other documents and practices that improve data exchange. \textit{Interest Groups} serve as a platform for communication and coordination among individuals, outside and within RDA, with shared interests in data sharing, exchange and interoperability. They produce surveys, recommendations, reports, and Working Group case statements. They should not be promoting specific projects or solutions, and they remain in operation as long as they remain active. There are currently (October 2015) about 15 Working Groups and 40 Interest Groups, with very different kinds of topics, from very technical ones to more sociological ones. The Working Groups established at the beginning of the organisation already produced their deliverables, which are displayed on the RDA web site. 

The first Working Groups were tackling very technical subjects. They were five which started at the beginning of the RDA, Data Foundation and Terminology, Data Type Registries, Metadata Standards Directory, Persistent Identifier Information Types, Practical Policies (which deals with machine actionable policies for collection management). Soon different kinds of topics were proposed, often following discussions in an Interest Group. Several disciplinary communities set up Working Groups. For instance, the agriculture community was present at the very beginning of the RDA. They seized the opportunity to discuss the interoperability aspects of the International Wheat Initiative\footnote{\url{http://www.wheatinitiative.org/}}, a major initiative launched at governmental level to coordinate global research on wheat, in this neutral, international forum. The Wheat Interoperability Working Group succeeded in producing a cookbook containing guidelines on how to describe, represent and share wheat data, a portal which references wheat related ontologies and vocabularies, and a prototype of a semantics knowledge base. Another example is more on construcing an important "sociological" building block of data sharing. The Repository Audit and Certification DSA-WDS Partnership Working Group is aligning two existic basic certification mechanism, those of the Data Seal of Approval\footnote{\url{http://datasealofapproval.org/en/}} and of the World Data System\footnote{\url{http://www.icsu-wds.org/}}, to build a common framework for certification of trusted repositories. A series of Groups set up in collaboration between the RDA and ICSU World Data System also deal with different aspects of data publication, including Publishing Data Workflows, Services and Bibliometric, with the participation of publishers.

The landscape of RDA Groups is now very diverse, with Groups focused on the basic technical infrastructure of data sharing, on data stewardship and services, on referencing and sharing, on community needs, and on domain science. Some groups are closer to the data providers' needs, others to the data users', with technical or sociological aims. The landscape is constantly evolving with new proposals, and currently several of the early Working Groups are discussing new Groups for follow-up activities.

The Technical Advisory Board follows the Group activities but does not supervise nor organise them: the RDA is really a bottom-up organisation. Groups are emerging to pool some aspects of the activities, for instance the Data Fabric Interest Group looks for common components and services to make the work along the data life cycle as efficient and reproducible as possible, including the outputs of the first Working Groups, and the work of several of the other Groups. The RDA/WDS Data Publishing Interest Group is another example: it serves as an umbrella for the cluster of Groups which deal with data publishing.

\section{Astronomy and the RDA}

Astronomy has been implementing data sharing for a long time now. FITS, although there are discussions about it, exists and serves as a basis. The ADASS conferences are a key yearly rendez-vous for all those who work on data pipelines and data sharing in the discipline. We managed to develop and network astronomical on-line resources very early after the advent of the internet. In addition the IVOA is sucessfully setting up a global discipline-wide framework for interoperability. 

Astronomy representatives were present in meetings from the early beginning of the RDA, but there was no urgent need to set up a disciplinary astronomy Interest Group, because the IVOA plays this role for us. The CDS has been a member of the series of European projects\footnote{\url{http://europe.rd-alliance.org/}} set up in support to the RDA. It is making sure to inject the lessons learnt from building the astronomical data framework and the IVOA in the construction of the RDA organisation and processes, which is on-going since it is still a young organisation.

The astronomy data infrastructure is not an isolated island. For for instance the IVOA made sure to use generic building blocks when possible, in particular for critical elements such as semantics and the registry of resources. The IVOA Vocabularies\footnote{\url{http://www.ivoa.net/documents/cover/Vocabularies.html}} are based on the W3C's Resource Description Framework (RDF) and Simple Knowledge Organization System (SKOS). As stated in the abstract of the standard, "by adopting a standard and simple format, the IVOA will permit different groups to create and maintain their own specialised vocabularies while letting the rest of the astronomical community access, use, and combine them. The use of current, open standards ensures that VO applications will be able to tap into resources of the growing semantic web." Similarly, the IVOA Registry of Resouces\footnote{\url{http://www.ivoa.net/documents/latest/RM.html}} is compliant with the OAI-PMH standard and includes the Dublin core, both being widely used in the digital library world. This made it relatively straightforward to include the IVOA Registry into the generic B2FIND registry of resources set by te European EUDAT\footnote{EUDAT \url{https://eudat.eu/} is the collaborative Pan-European infrastructure providing research data services, training and consultancy for researchers, research communities and research infrastructures and data centres.} project. 

The RDA is a unique gathering of people from across the world with different profiles involved in the sharing of scientific data. There are lots of useful ideas to catch from the work of the Working and Interest Groups. Also the RDA Plenaries gather hundreds of participants and are unique occasions for very rich informal interactions with people from other disciplines and with other backgrounds. This is particularly useful to share lessons learnt and discover interesting new methods.

Several Working Groups and Interest Groups are potentially of specific interest for the astronomy data providers. This is the case all those which work on data citation and Persistent Identifiers - the usage of DOIs (Digital Object Identifiers) is currently a hot topic for astronomy. Of particular interest for centres like CDS which build dynamic databases such as SIMBAD is the group working on (dynamic) data citation - their work aims at allowing users to cite a particular query to a database made at a particular time, with the capacity to retrieve the query result. Many topics are also more and more relevant in the current general landscape in which funding agencies more and more require the projects they fund to establish a Data Management Plan. One RDA Group works on Active Data Management Plans, and other groups also tackle critical building blocks, for instance certification of trusted digital data repositories. On a more focused technical topic, the RDA Research Data Provenance Interest Group is very interested in the work performed in collaboration between the IVOA and CTA to build a provenance framework adapted for the project. There are also "sociological" Groups in which we belong naturally, for instance Domain Repositories. Some astronomers are also interested in Data Rescue, and one co-chairs that RDA Group.

The advice to astronomy data providers is to have a look at the wealth of RDA Groups and discussions. The official outcomes are not so many for the moment, but their number is growing, and aside from the officially approved products of the Working Groups there are many results which can be of interest in our daily practice. We also tackle for our own needs many topics of interest for a wider community. We can improve our own methods and practices by participating actively in Group activities, and it will also be good for the Groups themselves to include our expertise. Finally, if you do not find a Group dealing with the subject you want to tackle, you can propose a Group with colleagues from other countries and regions also interested. If you do not have these international connections, you can build them by proposing a dedicated Bird of a Feather session for a RDA Plenary meeting. Interested people will join to discuss the topic and the possibility to propose a Group.

\acknowledgements Françoise Genova aknowledges support from the "RDA Europe - the European plug-in to the global Research Data Alliance (RDA)" project, funded by the European Commission under the 7th Framework Programme (GA632756).

\end{document}